# Spin torque and critical currents for magnetic vortex nano-oscillator in nanopillars

Konstantin Y. Guslienko[1,2*], Gloria R. Aranda[1], and Julian M. Gonzalez[1]

[1]*Dpto. Fisica de Materiales, Universidad del Pais Vasco, 20018 Donostia-San Sebastian, Spain*

[2]*IKERBASQUE, the Basque Foundation for Science, 48011 Bilbao, Spain*

We calculated the main dynamic parameters of the spin polarized current induced magnetic vortex oscillations in nanopillars, such as the range of current density, where a vortex steady oscillation state exists, the oscillation frequency and orbit radius. We accounted for both the non-linear vortex frequency and non-linear vortex damping. To describe the vortex excitations by the spin polarized current we used a generalized Thiele approach to motion of the vortex core as a collective coordinate. All the results are represented via the free layer sizes, saturation magnetization, Gilbert damping and the degree of the spin polarization of the fixed layer. Predictions of the developed model can be checked experimentally.

Key words: spin polarized current, magnetic nanopillar, nano-oscillators, magnetic vortex

[*]Corresponding author. Electronic mail: sckguslk@ehu.es

Now excitations of the microwave oscillations in magnetic nanopillars, nanocontacts and tunnel junctions by spin polarized current as well as the current induced domain wall motions in nanowires are perspective applications of spintronics.[1] A general theoretical approach to microwave generation in nanopillars/nanocontacts driven by spin-polarized current based on the universal model of an auto-oscillator with negative damping and nonlinear frequency shift was developed recently by Slavin and Tiberkevich [see Ref. 2 and references therein]. The model was applied to the case of a spin-torque oscillator (STO) excited in a uniformly magnetized free layer of nanopillar, and explains the main experimentally observed effects such as the power and frequency of the generated microwave signal. However, the low generated power ~ 1 nW of such STO prevents their practical applications. Recently extremely narrow linewidth of 0.3 MHz and relatively high generated power was detected for the magnetic vortex (strongly non-uniform state) nano-oscillators in nanopillars.[3] The considerable microwave power emission from a vortex STO in magnetic tunnel junctions was observed.[4] It was established that the permanent perpendicular to the plane (CPP) spin polarized current $I$ can excite vortex motion in free layer of the nanopillar if the current intensity exceeds some critical value, $I_{c1}$.[5] Then, in the interval $I_{c1} < I < I_{c2}$ ($I_{c2}$ is a second critical current) there is microwave generation at a frequency which smoothly increase with the current increasing.[6] This frequency corresponds to the vortex oscillations with a stationary orbit determined by the current value. If the current exceeds a critical value $I_{c2}$[7,8] the vortex steady state is not stable anymore, presumably because the vortex reaches a critical velocity[9] and reverses its core. The vortex with opposite core cannot be excited for the given current sign $I$ and the oscillations stop. Such excitation scheme is different from the current-in-plane (CIP) case, where one needs to apply a.c. CIP of about the resonance frequency to excite the vortex motion.[10] Low value of $I_{c1}$ and high value of $I_{c2}$ make the vortex CPP nano-oscillators attractive for applications as microwave devices. However, the calculations of the spin torque term (ST force) gave contradictory results for the ST force and $I_{c1}$. The standard STO approach of Ref. 2 is not applicable to the vortex dynamics due to specific damping term. The problem was reduced to the problem of vortex core reversal in the perpendicular to the layer plane magnetic field in Ref. 7 but vortex steady state dynamics was not accounted. Using the Thiele approach the ST force was calculated in Refs. 5, 8 which differs in 2 times from one calculated form the energy dissipation balance.[6] Non-linearity of the main governing parameters and the Oersted field of current were not accounted or accounted incorrectly. The critical current $I_{c2}$ has not yet been calculated.

In this Letter we present a simple and effective approach to calculate the ST force and the critical currents of the vortex STO in nanopillar. The approach is based on the Thiele formulation of the non-uniform magnetization dynamics[11] and conception of the linear spin excitations.[12] The nanopillar device consists of two ferromagnetic layers, typically FeNi or Co and a nonmagnetic metallic spacer, typically Cu, arranged in a vertical stack (Fig. 1). Magnetization of one layer is fixed (this layer is the so called polarizer), whereas the magnetization of the second layer of the nanopillar $\mathbf{M}(\mathbf{r},t)$ is free to rotate. The current of spin polarized electrons transfers some torque $\boldsymbol{\tau}_s$ from the polarizer, which excites magnetization dynamics of the free layer. We start from the Landau-Lifshitz equation of motion $\dot{\mathbf{m}} = -\gamma \mathbf{m} \times \mathbf{H}_{eff} + \alpha_{LLG} \mathbf{m} \times \dot{\mathbf{m}} + \gamma \boldsymbol{\tau}_s$, where $\mathbf{m} = \mathbf{M}/M_s$, $M_s$ is the saturation magnetization, $\gamma > 0$ is the gyromagnetic ratio, $\mathbf{H}_{eff}$ is the effective field, and $\alpha_{LLG}$ is the Gilbert damping. We use the ST term in the form suggested by Slonczewski,[13] $\boldsymbol{\tau}_s = \sigma J \mathbf{m} \times (\mathbf{m} \times \mathbf{P})$, where $\sigma = \hbar \eta / (2|e|LM_s)$, $\eta$ is the current spin polarization ($\eta=0.2$ for FeNi), $e$ is the electron charge, $L$ is the free layer (dot) thickness, $J$ is the current density, and $\mathbf{P} = P\mathbf{z}$ is the unit vector of the polarizer magnetization ($P=+1/-1$). We assume the positive vortex core polarization $p=+1$, $P=+1$ and define the current (flow of the positive charges) as positive $I>0$ when it flows from the polarizer to free layer. The spin polarized current can excite a vortex motion in the free layer if only $IpP > 0$ (only the electrons bringing a magnetic moment from the polarizer to free layer opposite to the core polarization can excite a vortex motion). Except $p$, the vortex is described by its core position in the free layer, **X**=(X,Y), and chirality $C=\pm 1$.[14] Let denote the Slonczewski's energy density which corresponds to the spin polarized current as $w_s$. Then, using the Thiele approach and the ST field $\partial w_s / \partial \mathbf{m} = a\mathbf{m} \times \mathbf{P}$, the ST force acting on the vortex in the free layer can be written as

$$F_{ST}^{\alpha} = -\frac{\partial}{\partial X_{\alpha}} \int dV w_s = aL\mathbf{P} \cdot \int d^2\boldsymbol{\rho} \left( \mathbf{m} \times \frac{\partial \mathbf{m}}{\partial X_{\alpha}} \right), \quad (1)$$

where $a = M_s \sigma J$, $\alpha = x, y$, $\boldsymbol{\rho} = (\rho, \varphi)$ is the in-plane radius vector, the derivative is taken with respect to the vortex core position **X** assuming an ansatz $\mathbf{m}(\boldsymbol{\rho},t) = \mathbf{m}[\boldsymbol{\rho}, \mathbf{X}(t)]$ (**m** dependence on thickness coordinate $z$ is neglected). $|\mathbf{X}|$ has sense of the amplitude of the vortex gyrotropic eigenmode.

We use representation of **m**-components by the spherical angles $\Theta, \Phi$ (Fig. 1) as $\mathbf{m} = (\sin\Theta\cos\Phi, \sin\Theta\sin\Phi, \cos\Theta)$ and find the expression for the ST force

$$\mathbf{F}_{ST} = aL \int d^2\boldsymbol{\rho} \sin^2\Theta \frac{\partial \Phi}{\partial \mathbf{X}}. \quad (2)$$

In the main approximation we use the decompositions $m_z(\boldsymbol{\rho}, \mathbf{X}) = \cos\Theta = m_z^0(\rho) + g(\rho)(\mathbf{X} \cdot \hat{\boldsymbol{\rho}})$, $\Phi(\boldsymbol{\rho}, \mathbf{X}) = \Phi_0 + m_0(\rho)[X\sin\varphi - Y\cos\varphi]$, where $m_z^0(\rho)$, $\Phi_0$ are the static vortex core profile and phase, $g(\rho) = 4pc^2\rho(1+\rho^2)/(c^2+\rho^2)^2$ is the excitation amplitude of the z-component of the vortex magnetization ($c = R_c/R$, $R_c$ is the vortex core radius, $\rho$, $|\mathbf{X}|$ are normalized to the dot radius $R$) and $m_0(\rho) = (1-\rho^2)/\rho$ is the gyrotropic mode profile.[12] One can conclude from Eq. (2) that only moving vortex core contribute to the ST force because the contribution of the main dot area where $\Theta = \pi/2$ is equal to zero due to vanishing integrals on azimuthal angle φ from the gradient of the vortex phase $\partial_X \Phi$ (it was checked accounting in $\Phi(\mathbf{\rho},\mathbf{X})$ the terms up to cubic terms in $X_\alpha$-components). This is a reason why the ST contribution is relatively small being comparable with the damping contribution.

The integration in Eq. (2) yields the ST force $\mathbf{F}_{ST} = \pi a L(\hat{\mathbf{z}} \times \mathbf{X})$. This force contributes to the Thiele's equation of motion $\mathbf{G} \times \dot{\mathbf{X}} = -\partial_\mathbf{X} W + \hat{D}\dot{\mathbf{X}} + \mathbf{F}_{ST}$, where $\mathbf{G} = \hat{\mathbf{z}} 2\pi p L M_s / \gamma$ is the gyrovector, $\hat{D}$ is the damping tensor. The vortex energy $W(\mathbf{X})$ and restoring force $\mathbf{F}_R = -\partial_\mathbf{X} W$ can be calculated from an appropriate model[14] (the force balance is shown in Fig. 2). For circular steady state vortex core motion the $\dot{\mathbf{X}} = \boldsymbol{\omega} \times \mathbf{X}$ relation holds, which allows calculating $J_{c1}$. To calculate the vortex steady orbit radius $R_s = |\mathbf{X}|$ we need, however, to account non-linear on $X_\alpha$ terms in the vortex damping and frequency (the account only non-linear frequency as in Ref. 5 is not sufficient). The gyrovector also depends on $\mathbf{X}$, but this dependence is essential only for the vortex core $p$ reversal, where $\mathbf{G}$ changes its sign. As we show below, the most important non-linearity comes from the damping tensor defined as

$$D_{\alpha\beta}(\mathbf{X}) = -\alpha_{LLG} \frac{M_s}{\gamma} \int dV \frac{\partial \mathbf{m}}{\partial X_\alpha} \cdot \frac{\partial \mathbf{m}}{\partial X_\beta}, \qquad (3)$$

or $D_{\alpha\beta} = -\alpha_{LLG}(M_s L/\gamma) \int d^2\boldsymbol{\rho} \left[\partial_\alpha \Theta \partial_\beta \Theta + \sin^2 \Theta \partial_\alpha \Phi \partial_\beta \Phi\right]$ in $\Theta, \Phi$-representation Accounting $D_{\alpha\beta} = D\delta_{\alpha\beta}$ and introducing dimensionless damping parameter $d = -D/|G| > 0$[15] we can write the equation for a steady state vortex motion with the orbit radius $R_s = |\mathbf{X}|$: $d(s)|G|\omega_G(s)sR = F_{ST}^\varphi$ from which $s = R_s / R$ and the critical currents $J_{c1}$, $J_{c2}$ can be found. In the second order non-linear approximation $d(s) = d_0 + d_1 s^2$, $\omega_G(s) = \omega_0 + \omega_1 s^2$ and $F_{ST}^\varphi = \pi a L R s$, where $\omega_G$ is the vortex precession frequency, $\omega_1 > 0$ is a function of the dot aspect ratio $\beta = L/R$ calculated from the vortex energy decomposition $W(s)$ in series of $s = |\mathbf{X}|/R$. It can be shown that $\omega_0(\beta) = (20/9)\gamma M_s \beta[1 - 4\beta/3]$ and $\omega_1(\beta) \approx 4\omega_0(\beta)$ for quite wide range of $\beta = 0.01$-$0.2$ of practical interest, whereas considerably larger non-linearity $\omega_1(\beta)/\omega_0(\beta) = 42.8$ was calculated in Ref. 5 due to incorrect account of the magnetostatic energy. We use the pole free model of the shifted vortex $\mathbf{m}[\boldsymbol{\rho}, \mathbf{X}(t)]$, where the dynamic magnetization satisfies the strong pinning boundary condition at the dot circumference[16] $\rho = R$. The damping parameters are $d_0 = \alpha_{LLG}(5/8 + \ln(R/R_c)/2)$, $d_1 = \alpha_{LLG}(R^2/R_c^2 - 8/3)/4$. We need also to account for the Oersted field of the current, which leads to contribution to the vortex frequency proportional to the current density $\omega_0 = \omega_0(\beta, J) = \omega_0(\beta) + \omega_e J$, where $\omega_e = (8\pi/15)(\gamma R/c)\xi C$, $\xi = 1 + 15(\ln 2 - 1/2)R_c/8R$ is the correction for the finite core radius $R_c \ll R$. In the linear approximation we get the equality of the damping force and the ST force (negative damping) as the condition to find the value of $J_{c1}$, with the contribution of the Oersted field of the current accounted. The values of $F_{ST}$, $J_{c1}$ coincide with calculation of Ref. 6 conducted by the method of the damping energy balance and differ by the multiplayer 2 from Refs. 5, 8. The first critical current is $J_{c1} = d_0\omega_0/(\gamma\sigma/2 - d_0\omega_e)$ and the steady state vortex orbit radius is

$$s(J) = \lambda\sqrt{J/J_{c1} - 1}, \quad J > J_{c1}, \quad \lambda^2 = \frac{1}{2}\frac{\gamma\sigma J_{c1}}{[d_1\omega_0(J_{c1}) + d_0\omega_1]} \quad (4)$$

In this approximation the vortex trajectory radius $s(J)$ increases as square root of the current overcriticality $(J - J_{c1})/J_{c1}$ (for the typical parameters and R=80-120 nm we get λ=0.25-0.30) and the vortex frequency $\omega_G = \omega_0(\beta) + \omega_e J + \lambda^2(J/J_{c1} - 1)\omega_1$ increases linearly with J increasing. The vortex steady orbit can exist until the moving vortex crosses the dot border s=1 or its velocity $|\dot{\mathbf{X}}|$ reaches the critical velocity $v_c$ defined in Ref. 9. The later allows to write equation for the second critical current $J_{c2}$ as $\omega_G(J)s(J)R = v_c$. Substituting to this expression the equations for $\omega_G(J)$ and $s(J)$ derived above we get a cubic equation for $J_{c2}$ in the form $[\gamma\sigma J_{c1}/2d_0 + (\omega_e J_{c1} + \omega_1 \lambda^2)x]x^{1/2} = v_c/\lambda R$, $x = (J_{c2}/J_{c1} - 1)$. This equation has one positive root $x_c$ and the value of $J_{c2}$ can be easily calculated (Fig. 3). The former condition ($s=1$) gives the second critical current $J'_{c2} = (1 + 1/\lambda^2)J_{c1}$. More detailed analysis shows that both the mechanisms of the high current instability of the vortex motion are possible depending on the dot sizes $L$, $R$, and the critical current is the lower value of the currents $J_{c2}$, $J'_{c2}$. The vortex core reversal inside the dot occurs for large enough $R$ (> 100 nm) and $L$. For the typical sizes $L=10$ nm, $R=120$ nm and $C=1$, the critical currents are $J_{c1}=6.3 \times 10^6$ A/cm$^2$ ($I_{c1}=2.9$ mA), $J_{c2}=1.13 \times 10^8$ A/cm$^2$ ($I_{c2}=51$ mA), and for $L=5$ nm, $R=100$ nm we get $J_{c1}=1.8 \times 10^6$ A/cm$^2$ ($I_{c1}=0.56$ mA), $J'_{c2}=2.7 \times 10^7$ A/cm$^2$ ($I'_{c2}=8.4$ mA).

In summary, we calculated the main physical parameters of the spin polarized CPP current induced vortex oscillations in nanopillars, such as the critical current densities $J_{c1}$, $J_{c2}$, the vortex steady state oscillations frequency and orbit radius. All the results are represented via the free layer sizes ($L$, $R$), saturation magnetization, Gilbert damping and the degree of the spin polarization of the fixed layer. These parameters can be obtained from independent experiments. We demonstrated that the generalized Thiele approach is applicable to the problem of the vortex STO excitations by the CPP spin polarized current. The spin transfer torque force is related to the vortex core only.

The authors thank J. Grollier and A.K. Khvalkovskiy for fruitful discussions. K.G. and G.R.A. acknowledge support by IKERBASQUE (the Basque Science Foundation) and by the Program JAE-doc of the CSIC (Spain), respectively. The authors thank UPV/EHU (SGIker Arina) and DIPC for computation tools. The work was partially supported by the SAIOTEK grant S-PC09UN03.

**Captions to the Figures**

Fig. 1. Sketch of the magnetic nanopillar with the coordinate system used. The upper (free) layer is in the vortex state with non-uniform magnetization distribution. The polarizer layer (red color) is in uniform magnetization state with the magnetization along $Oz$ axis. The positive current $I$ (vertical arrow) flows from the polarizer to free layer.

Fig. 2. Top view of the free layer with the moving vortex. The arrows denote the force balance for the vortex core. The spin torque ($\mathbf{F}_{ST}$), damping ($\mathbf{F}_D$), restoring ($\mathbf{F}_R$) and gyro- ($\mathbf{F}_G$) forces are defined in the text. The vortex core steady trajectory $R_s$ is marked by orange color. The vortex chirality is $C=+1$.

Fig. 3. Dependence of the critical currents $J_{c1}$ (solid red line), $J_{c2}$ (dashed green line) and $J'_{c2}$ of the vortex motion instability on the radius $R$ of the free layer. $L=$ 10 nm, $M_s =$800 G, $\eta =$0.2, $\alpha_{LLG} = 0.01$, $\gamma/2 =$2.95 MHz/Oe, $R_c=$12 nm. The vortex STO motion is stable at $J_{c1} < J < \min(J_{c2}, J'_{c2})$.

Fig. 1.

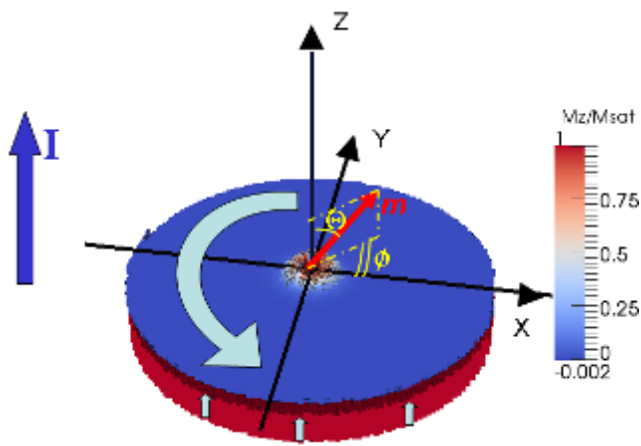

Fig. 2.

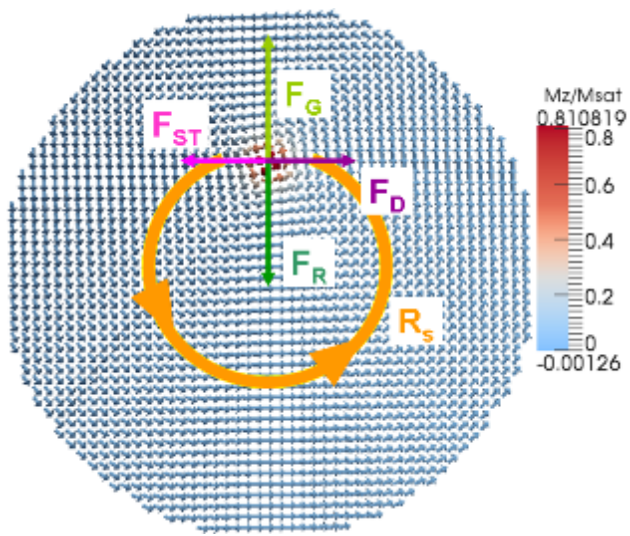

Fig. 3.

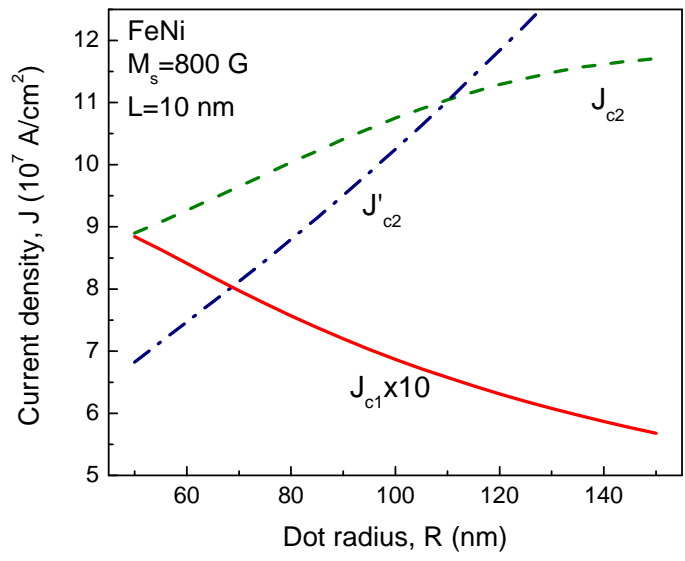